\magnification=\magstep1
\font\Large=cmr14
\font\large=cmr12
\font\boldmath=cmmib10
\def\be{{\bf e}}
\def\bomega{\hbox{\boldmath\char'041}}
\def\br{{\bf r}}
\def\gtorder{\mathrel{\raise.3ex\hbox{$>$}\mkern-14mu
             \lower0.6ex\hbox{$\sim$}}}
\def\ltorder{\mathrel{\raise.3ex\hbox{$<$}\mkern-14mu
             \lower0.6ex\hbox{$\sim$}}}
\def\br{{\bf r}}
\def\ecrit{E_{\rm crit}}
\def\etal{{\sl et al}\ }
\centerline{{\Large The Time Scale of Escape from Star Clusters}}
\bigskip\bigskip
\centerline{{\large T. Fukushige}}
\smallskip
\centerline{Department of General Systems Studies,}
\centerline{ College of Arts and Sciences,}
\centerline{ University of Tokyo, }
\centerline{3-8-1                Komaba,}
\centerline{ Meguro-ku,}
\centerline{ Tokyo 153-8902,}
\centerline{ Japan}
\bigskip
\centerline{{\large D.C. Heggie}}
\smallskip
\centerline{Department of Mathematics and Statistics,}
\centerline{ University of Edinburgh,}
\centerline{King's Buildings,}
\centerline{Edinburgh EH10 5NL,}
\centerline{UK}
\bigskip
\noindent{\bf Abstract}

\medskip

In this paper a cluster is modelled as a smooth potential (due to the
cluster stars) plus the steady tidal field of the Galaxy.  In this
model there is a minimum energy below which stars cannot escape.
Above this energy, however, the time scale on which stars escape
varies with the orbital parameters of the star (mainly its energy) 
in a way which we attempt to quantify, with both
theoretical arguments and computer simulations.  Within the
limitations of the model we show that the time scale is long enough to
complicate the interpretation of full $N$-body simulations of
clusters, and that stars above the escape energy may remain bound to
the cluster for about a Hubble time.


\bigskip

\noindent {\bf 1. Introduction}

\medskip

Our initial reason for studying this problem stemmed from a
collaborative study of star cluster evolution (Heggie \etal 1998), the
aim of which was to compare results obtained with different codes and
models for the same initial conditions.  These initial conditions
corresponded to a cluster with about $2.5\times10^5$ stars moving in a
circular orbit about a point-mass galaxy.  Among the resulting
conclusions was a noticeable problem with scaling the results of
$N$-body simulations.  ($N$-body simulations use smaller $N$ than the
number of stars in the cluster under study, and so it is necessary to
scale the results appropriately.)  When this was done on the basis of
the theory of relaxation (e.g. Spitzer 1987), the results appeared to
depend significantly on $N$, in the sense that the larger models
(i.e. larger $N$) lost mass more quickly (when scaled to the cluster
under study) than smaller models.  This directed our attention to the
escape rate from tidally bound clusters with unequal stellar masses.

It turns out that the time scale on which stars escape is one half of
the explanation for the scaling problem with the $N$-body models.  The
other half of the explanation is the manner in which the initial
conditions were set up, which leads to an initial population of stars
with energies above the energy of escape (``primordial escapers'').  Though all three
of the  $N$-body codes that were used in the collaborative experiment 
were written independently, all three
constructed initial models which were inconsistent with the tidal
field.  In these problems the gravitational field is given by
$$
\phi=\phi_{\rm c} + {1\over 2}\omega^2(z^2 - 3x^2),\eqno(1)
$$
where $\phi_{\rm c}$ is the field due to the cluster stars, and the
remainder is a combination of tidal and centrifugal terms (cf. Spitzer
1987).  In this
equation $\omega$ is the
angular velocity of the cluster round the galaxy, and the coordinates
are defined in \S2.1.
In the $N$-body simulations initial
conditions were set up by first constructing  an {\sl isolated}
model, and then immersing  this in
the tidal field.  Unfortunately this creates an initial model in which
many stars have energy above the escape energy (Fig.1).

This flaw would not matter if these stars with excess energy would
escape quickly.  It would be possible to wait until they had mostly
disappeared, and then examine the relaxation-driven evolution of those
that remain.  Unfortunately, however, it turns out that the time scale on which these
stars escape may be very long, and some do not escape at all (if we
take the cluster potential to be fixed).  The manner in which the
distribution of escape times depends on $N$ is not known, but it
evidently complicates the scaling of the models.  It was our attempt
to understand this problem which motivated our investigation.

Many previous investigations have been devoted to the escape rate of stars from
clusters, even if we restrict discussion to those with tidal boundary
conditions.  Some theoretical investigations (e.g. Chandrasekhar 1943,
King 1960, Kwast 1977, Pietrovskaia 1977, Johnstone 1993) actually
compute the rate at which stars cross the threshold of {\sl energy} for escape,
and it is assumed implicitly that the escape itself is immediate.
This remark applies also to estimates from Fokker-Planck models
(e.g. Spitzer \& Shull 1975, Lee \& Goodman 1995), with some
exceptions (Lee \& Ostriker 1987, Takahashi \& Portegies Zwart 1998).  Similarly,
$N$-body models have been used to determine escape rates (e.g.  Wielen
1975, de la Fuente Marcos 1995) but usually no distinction is made
between the time taken to raise the energy of a star above the escape
threshold and the time taken to cross the tidal radius.

More apposite to our interests are the simulations of Chernoff \&
Weinberg (1991), who observed the rate of escape of stars from a
cluster in a tidal
field, with initial conditions which, like ours, include stars with
energy above the escape energy.  They found that almost 15\% of all
the stars escaped beyond a distance of about 3 tidal radii within 100
of their time units, which corresponds to about 500 crossing times,
and that stars on prograde orbits escaped slightly faster.  Escape is
particularly rapid within the first 100 crossing times.  This
corresponds to roughly the core collapse time of some of the models in
the collaborative study already mentioned, which suggests that the
continued escape of stars is a significant complication.  Chernoff \&
Weinberg also measured the rate of loss of particles as a function of
energy, a point to which we devote considerable attention in our
paper.  What complicates the interpretation of their results for our
purposes, however, is the fact that many stars were initially outside
the tidal radius, and also we are interested in escape across the
tidal radius, rather than some larger radius.


The  simulations of Baumgardt (1997) and Johnston, Sigurdsson \& Hernquist (1999)  take this kind of study
further by placing the cluster on an eccentric galactic orbit.  In the
present paper, however, we restrict attention to the circular case,
with which the collaborative experiment was also concerned.

Our aim  is to study the distribution of escape
times, as a function of energy, for stars initially located inside
the tidal radius.  In addition to the results of new numerical
simulations not unlike those of Chernoff \& Weinberg (Section 2), we
provide some information on how these results depend on the initial
cluster model.  We also show that it is possible to give some
theoretical estimates of the escape time scale and its energy
dependence (Section 3), which are compared with our numerical results
in Section 4.  Section 5 summarises our conclusions and applies them
to a number of results in the literature.

\bigskip\noindent
{\bf 2.  Numerical Results}

\medskip\noindent
{\bf 2.1 Method }

We calculated the orbits of stars in a smoothed potential due to cluster
stars plus the steady tidal field of the Galaxy, in order to obtain the
escape time from the cluster.  We set the initial center of mass of the
cluster at the origin $(x,y,z)=(0,0,0)$ with axes oriented so that the
position of the galactic center is $(-R_{\rm g},0,0)$.  We assumed that
the size of the cluster is much smaller than $R_{\rm g}$.  If the cluster
revolves around the galactic centre at constant angular velocity
$\bomega=(0,0,\omega)$, the equation of motion of a star can be
expressed as
$$
{{\rm d}^{2}{\br}\over {\rm d}t^2} =  
-\nabla \phi_{\rm c}(r) 
- 2\bomega\times{{\rm d}{\br}\over {\rm d}t} 
+ \omega^2(3x{\be}_x-z{\be}_z),    \eqno(2)
$$
where ${\br}$ is the position of the star, and ${\be}_x$, ${\be}_z$ are unit vectors that point along the $x$ and $z$ axes,
respectively. 

The first term on the right-hand side in eq.(2) is the
gravitational acceleration from the cluster.  We used King's models
(King 1966) for the potential of the cluster, and the potential is
smooth and fixed.  We used standard units such that $M=G=-4E_{\rm
c}=1$, where $M$ is the total mass of the cluster, $G$ is the
gravitational constant, and $E_{\rm c}$ is the total self-energy of the
cluster (Heggie and Mathieu 1986).  We performed calculations with 
King models which differ in the dimensionless central potential,
$W_0= 3, 7, 12$.  In order to compute the acceleration of the King
model, we used cubic spline interpolation between data obtained by
numerical solution (e.g.  Press et al.  1992). 

The second term on the right-hand side in eq.(2) is the Coriolis
acceleration, and the third term is a combination of the centrifugal and
tidal forces.  Here we assumed that the cluster moves on a circular
orbit in a spherically symmetric galactic potential, taken to be that of
a distant point mass.  The angular velocity, $\omega$, is given by
$$
\omega = \sqrt{GM\over 3r_{\rm t}^3}\eqno(3)
$$
where $r_{\rm t}$ is the tidal radius of the cluster.  We set $r_{\rm
t}$ to be equal to the tidal radius of the King model, and so obtained
$\omega$. 

We calculated orbits of 1000 stars to obtain the distribution function
of escape time, for several values of the dimensionless energy excess, ${\tilde
E}=(E-E_{\rm crit})/\vert E_{\rm crit}\vert$, where $E$ is the energy
of a star of speed $v$, 
given by
$$
E={v^2\over 2}+\phi_{\rm c}+{1\over 2}\omega^2(z^2-3x^2),
$$
and $E_{\rm crit}$ is the potential at the Lagrangian points, given by
$$
E_{\rm crit} = -3GM/2r_{\rm t}.\eqno(4)
$$  
(Fig.2).    The criterion of escape is simply geometric: escapers are defined
to be those stars beyond the tidal radius. 

We performed calculations of orbits from two kinds of initial spatial
distribution of stars: a ``uniform'' distribution, and a ``King" or
``non-uniform" distribution.  In the former distribution, the
positions of stars are distributed uniformly inside the tidal radius
of the cluster.  In the latter distribution, the positions of the
stars are distributed in the same way as the primordial escapers in a
King model immersed in the tidal field.  In this case the stars tend
to be distributed in the inner regions of the cluster due to the
decrease of the potential by the tidal field in the outer region (see
Fig.1).  In either case the velocity is distributed isotropically,
and its magnitude is determined
by the given value of ${\tilde E}$.

We integrated the orbits of the stars by means of a fourth-order
Hermite integration scheme (Makino \& Aarseth 1992) with a variable
time step algorithm.  The integration error in energy was about $5
\times 10^{-3}$ at most, and typically of order $10^{-3}$ at the point
where the longest simulation was stopped ($t=8\times10^5$, $W_0=3$,
$\tilde E=0.03$). The error varied almost linearly with time.  Conversely,
the point at which integrations were stopped, and the smallest value
of $\tilde E$, were determined from the maximum error.


\medskip\noindent
{\bf 2.2 Results}
\smallskip

A typical result is given in Fig.3.  It shows the distribution of
escape times from a King model with $W_0=3$, when the initial
conditions are selected as in the collaborative experiment (i.e. a
``King'' distribution, cf.\S2.1).  Results
are presented for different values of the stellar energy, in terms of
$\tilde E$.

The integrations were terminated at some large time, corresponding to
the vertical lines at the right side.  Even so, it is clear that there
is a significant fraction of initial conditions for which the escape
time is effectively infinite.  This fraction decreases with increasing
energy excess $\tilde E$, and varies somewhat from one model to
another (Fig.4).  In the case of non-uniform initial conditions with 
$W_0 = 3$ (i.e. the dashed line with triangles) the decrease is
perhaps distorted by the fact that the termination times of the runs
with different $\tilde E$ are different.  
Empirically, the fraction of non-escapers is about
$f_{\rm n} \simeq 0.17 - 0.6\tilde E$ for uniform initial conditions and about
$0.26 - 0.9\tilde E$ for non-uniform conditions.    Further inspection of our results showed that a
majority of these non-escapers are in retrograde orbits initially,
i.e. the sense of their motion around the cluster is the same as the
sense of motion of the cluster about the galaxy.

Now we turn attention to the particles which escaped within the time
limit of the simulations.  The escape time decreases with increasing
$\tilde E$.  For the example shown in Fig.3 the median escape time
(i.e. $P = 0.5$) decreases by approximately two orders of magnitude as 
$\tilde E$ increases from $0.03$ to $0.24$.  For a given $\tilde E$
the escape time is slightly smaller for uniform than for non-uniform
initial conditions.  Presumably part of the reason for this is that
a larger fraction of initial conditions lie at large radii for uniform 
initial conditions.  For non-uniform initial conditions the escape
time decreases slightly with increasing $W_0$, presumably for the same 
reason, at least in part.  This is not a complete explanation,
however, as there is also a decrease of escape time with increasing
$W_0$ when the initial conditions are sampled uniformly.



We postpone further discussion of these results until some theoretical 
considerations have been developed in the following section.

\bigskip\noindent
{\bf 3. Theory}

\medskip\noindent
{\bf 3.1 Overall escape rate}
\smallskip

The Lagrange points are saddle points of the effective potential, and when the energy $E$ is just
above $\ecrit$, a star must pass close to one of these saddles in order
to escape.  It is not difficult to compute an upper bound on the rate
at which phase volume crosses each of the surfaces $x=\pm r_{\rm t}$ in
phase space, and
hence an upper bound on the escape rate follows if we know the phase
density of stars.  The calculation is a simple case of a more
general theoretical result about flow near saddles (MacKay 1990), but
the proof we give is more pedestrian.

Shifting the origin to a saddle point and expanding to second order,
we find that 
$$
\phi - \ecrit={1\over2}\omega^2(-9x^2 + 3y^2 + 4z^2).\eqno(5)
$$
The one-way flux of phase volume, per unit energy, across $x=0$ is simply
$$
{\cal F}=\int_{\dot x>0}\delta(\phi + {1\over2}(\dot x^2 + \dot y^2 +
\dot z^2) - E) \dot x {\rm d}\dot x {\rm d}\dot y {\rm d}\dot z {\rm d}y {\rm d}z.\eqno(6)
$$
where $\phi$ is evaluated at $x=0$.  The integrations are readily 
done, but we must double the result as there are two saddles
past which stars can escape.  The final answer is thus
$$
{\cal F}={2\pi^2(E-\ecrit)^2\over\sqrt{3}\omega^2}.\eqno(7)
$$

A familiar calculation shows that the phase space volume per unit
energy is
${\cal V}=4\pi\int \sqrt{2(E-\phi)} {\rm d}^3\br$ over the appropriate
domain in space.  This does not depend sensitively on $E$ in the
vicinity of $E=\ecrit$, and so we evaluate it there.  The
integration may be estimated by a Monte Carlo technique, but in this
section  we
approximate $\phi_{\rm c}$ with the potential of a point mass, and then
the integrand is mildly unbounded.  This can be handled by writing the
integrand as $\sqrt{2(E-\phi)r}r^{-1/2}$, where $r^2=\br.\br$, and 
sampling points with density proportional to $r^{-1/2}$.  Our result
is 
$$
{\cal V}=4\pi C\sqrt{2}(GM)^{4/3}\omega^{-5/3},\eqno(8)
$$ 
where $C=0.40\pm0.01$. 
It follows that the time scale for escape of phase volume is 
$$
t_{\rm e}={{\cal V}\over{\cal F}} =
{2C\sqrt{6}\over\pi}{(GM)^{4/3}\omega^{1/3}\over
(E-\ecrit)^2},\eqno(9)
$$
or, in dimensionless form, $\omega t_{\rm e} = \displaystyle{{2^{7/2}C\over
3^{13/6}\pi\tilde E^2}}$.

In order to turn this result into an estimate of the time scale of the
escape of stars, there are two issues to be addressed.   One is
the distribution of stars within phase space, the effect of which
was discussed from a numerical point of view in Sec.2.2.  The second is the fact that eq.(7) is
strictly only an upper limit on the escape rate, which we consider
qualitatively in the next section.

It is worth noting that there is an appealing ``physical'' argument
for the escape rate which nevertheless leads to a wrong answer.  In
the coordinates of eq.(5), it is clear that an escaping star must
pass through an elliptical aperture at $x=0$ with semi-axes
$\sqrt{2(E-E_{\rm crit})/3}/\omega$ and $\sqrt{(E-E_{\rm crit})/2}/\omega$,
respectively.  There are two such apertures, and their total area is a
fraction $\displaystyle{{E-E_{\rm crit}\over2\sqrt{3}\omega^2r_{\rm t}^2}}$ of
the area of a sphere of radius $r_{\rm t}$.  If we assume that this fraction
of the stars find their way through one of these apertures in each
crossing time, which is of order $2\pi/\omega$, we find that the time
scale of escape varies as $(E-E_{\rm crit})^{-1}$, in contrast with
eq.(9).  Presumably what is missing is the condition that the star should
approach the aperture from the correct direction.

Actually, there is a correct physical
interpretation of these results, though not an intuitive one.  In the
two-dimensional problem of motions in the plane $z=0$, we see from
eq.(6) that 
$$
{\cal F} = \oint \dot y {\rm d}y,\eqno(10)
$$ 
where $\phi + \dot y^2/2 =
E$.  This expression for ${\cal F}$ resembles an action integral.  In
fact it is the action of an unstable periodic orbit situated close to
the $L_1$ (saddle) point (cf.\S3.2.2).  We are not aware, however, of any
comparable interpretation of the three-dimensional result, i.e. eq.(7).

\medskip\noindent {\bf 3.2 Qualitative Results on Escape}

\smallskip\noindent
3.2.1 {\sl Non-Escapers}

Let us consider the case of motion in the $x,y$ plane, and again take
$\phi_{\rm c}=-GM/r$.  We change to units in which $GM = \omega = 1$.  Fig.5
shows the surface of section $y=0$ (with coordinates $x, \dot x$) for
motions of energy $\ecrit$, and the curve $\dot y=0$.  More than
half of the surface is occupied by apparently regular orbits when
$E=\ecrit$, and the proportion does not decrease dramatically if $E$
exceeds $\ecrit$ slightly.  Such stars would not escape.  It is not
clear whether the same conclusion applies in the three-dimensional
problem, because of Arnold diffusion.  Nevertheless it is plausible
that these apparently regular motions correspond approximately to
the stars which do not escape in the numerical simulations of
Sec.2.2.  The presence of such stars implies that eq.(9) overestimates
the escape time scale, because eq.(8) overestimates the volume of
phase space that can escape.

Incidentally, the regular orbits correspond mainly to retrograde motions,
i.e. motions in which the star revolves around the cluster in the
opposite sense to the motion of the cluster around the galaxy.  The
stability of such motions is well known in analogous problems
(e.g. Benest 1974, Keenan \& Innanen 1975, Huang \&
Innanen 1983, Ross, Mennim \& Heggie
 1997).  The range of such orbits has been studied numerically for
the planar problem with a point-mass potential by H\'enon (1970) and Brunini(1996), and for a
cluster-like potential by Jefferys (1976).  He found that the measure of
permanently bound orbits decreases with increasing energy,
and is negligibly small for $\tilde E\gtorder0.3$.

\smallskip\noindent
3.2.2 {\sl Escapers}


So far we have seen that stars with a given energy above the escape energy
appear to fall into two kinds, depending on whether or not they
eventually escape.  Now we consider those that do eventually escape.
As shown long ago by Hayli (1970), it 
is clear that such stars must closely approach one or other Lagrange point, and
pass the narrow neck which opens out in the surface of section of
Fig.5 when $E>E_{\rm crit}$.  It will be useful to consider some
features of the orbits within this region.  We concentrate on motions
in the $x, y$ plane, but qualitatively similar conclusions follow for
three-dimensional motions.

The Lagrange points exist precisely at
$E=E_{\rm crit}$, but at $E>E_{\rm crit}$ there is 
an unstable periodic orbit situated close to each of the these points
(cf. H\'enon 1969 and the Appendix).  This
orbit shrinks to
the unstable equilibrium point as $E\to E_{\rm crit}+$. 
In the present context it exerts a strong controlling
influence on would-be escapers, as we shall see.

Precisely at $E_{\rm crit}$
there is only one orbit that approaches the equilibrium point as
$t\to\infty$ (Fig.6).
At each slightly higher energy there is a one-parameter family
of orbits approaching the periodic orbit (at different ``phases'').  
Fig 7 shows one of
these.  In phase space the set of all such orbits (at fixed energy)
forms a tubular surface\footnote{$^\ast$}{In dynamical systems theory
this is called the {\sl stable invariant manifold} of the periodic
orbit}.  Fig.8 shows the final section of several such orbits,
and there are
enough to allow the tube (or at least its projection onto the $x$, $y$
plane) to be visualised.  By examining the
equations of motion in the vicinity of the Lagrangian point, it is not
hard to see that it is orbits {\sl inside} this tube which lead to escape.
Indeed the flux of phase space along this tube is equal to the value
in eq.(10).  This tube is rather like a hose
connecting the interior of the cluster (at least, the section in the
$x, y$ plane) with the outside world.

If we follow backwards the tube of orbits leading to the periodic
orbit we see that it must approximately follow the orbit shown in
Fig.6.  It lies roughly within the region of the surface of section
which, at energy $E_{\rm crit}$ (Fig.5), is occupied by chaotic orbits.
It is reasonable to conjecture that the tube, as it is followed
backwards, becomes increasingly distorted and convoluted within this
region of apparently chaotic motions.  Consideration of phase space
volume immediately shows, however, that the tube cannot remain in this region
indefinitely.  Increasingly it must connect with the region outside
the cluster, corresponding to motions which are temporarily captured
within $r_{\rm t}$.  Indeed the time scale on which this happens must be of
order $t_{\rm e}$, given by eq.(9) for the three-dimensional problem.  Returning to the direction in which
time increases, we see that stars can enter the cluster from outside,
remain within the cluster (inside the tube) for some time, and then
escape again.

\smallskip\noindent
3.2.3 {\sl The Distribution of Escape Times}

Now we can construct a qualitative picture of the rate at which stars
escape.   Consider an experiment in which stars are distributed within
the cluster at some energy above $E_{\rm crit}$.  (The numerical results
of such experiments are described in Sec.2.)  Some of these stars will
lie in the regular region of phase space (Fig.5), and never escape.
Others will lie within the tube of escapers, and will initially escape
at a rate given by eq.(7), multiplied by their density in phase
space.  On a time scale of order $t_{\rm e}$, however, the rate of escape
will diminish, as the phase space which escapes consists increasingly
of parts of phase space which have been temporarily captured from
outside.

This picture can be given considerable precision in the language of
chaotic transport (Wiggins 1992), though it is beyond the scope of
this paper.  This theory, combined with appropriate numerical results,
permits a computation of the manner in which the rate of escape
diminishes with time, which in turn is determined by the distribution
of escape times.  In some model problems the distribution of escape
times is approximately exponential, and in others it is approximately
a power law.  The division into non-escapers and escapers, with a
power-law distribution of escape times for the latter, is also found
in numerical studies of other idealised escape problems (Kandrup \etal
1999). 

There is one situation where we may estimate the distribution of
escape times rather easily.  Consider first an orbit on the surface of
the tube which
approaches the Liapounov orbit as $t\to\infty$: its escape time is
infinite.  An orbit just inside the surface of the tube spends a long
time close to the Liapounov orbit, before eventually escaping.  In
fact, from familiar estimates of flow near a saddle point, it is easy
to see that the escape time varies nearly as $\ln(1/d)$, where $d$ is
the distance of the orbit from the surface of the tube.  This leads to
an exponential distribution of escape times.  Though it seems that
this argument applies only to orbits lying close to the tube, we have
argued that this tube is highly convoluted throughout a large region
of phase space, which would imply that the resulting distribution of
escape times is more widely applicable.

The problem of the distribution of escape times is actually a good
deal more complicated.  There are in
fact {\sl two} places (corresponding to the two Lagrangian points)
where stars may escape from the cluster.  Each has its own ``tube'' of
orbits with infinite escape times, though it is not clear to what
extent these two tubes are intertwined with each other.  Further
complexity is added by motions which are asymptotic to other periodic
orbits inside the cluster.  Some of these orbits and the associated
complexity may be glimpsed in the numerical calculations of Murison
(1989), who computed the {\sl capture} times of a large number of
orbits in a similar problem.  By time-reversibility, however,
similar conclusions can be reached about escape
times.

In other dynamical problems (e.g. the breakup of triple systems,
cf. Anosova \& Orlov 1994) the rate of escape diminishes
approximately exponentially with time.  Despite the complications,
then, we may at least {\sl conjecture} similar behaviour in the
present problem, on the basis of little more than the above simplified
arguments and analogies.  The empirical evidence will be considered
in the next section.

\bigskip\noindent {\bf 4. Interpretation of Numerical Results}


We first consider the predicted dependence on $\tilde E$ of the escape time
scale, i.e. $t_{\rm e}\propto \tilde E^{-2}$ (eq.[9]), taking the data of
Fig.3 as an example.  If we exclude the particles which have not
escaped at the end of each set of computations, and scale the escape
time to 
$$
\tilde t = \omega t\tilde E^2,\eqno(11)
$$ 
the result is Fig.9.  After an early transient the approximate
coincidence of the curves is a measure of the success of the predicted 
energy dependence.

In order to analyse this data we fitted a model based loosely on the
theory of \S3.  If we assume that stars escape on a time scale $t_{\rm e}$
then 
$$
P(t) = \exp(-t/t_{\rm e})\eqno(12)
$$ 
and $P(\tilde t) = \exp(-\tilde t/\tilde
t_{\rm e})$, where $\tilde t_{\rm e}$ is defined in the obvious way.   If $\tilde
t_{\rm e}$ is taken from eq.(9), the result is plotted as the dashed line
in Fig.9.  As can be seen this is a hopelessly poor fit:  the time
scale is too short by about 1 dex.  Recall, however, that this theory
was
derived from arguments of escape of phase space from the field of a
point mass, whereas phase space is sampled non-uniformly in the
numerical data and the potential is far from that of a point mass.
Both factors compare better if we turn to high-concentration models
and a uniform distribution of initial conditions, and then the
predicted time scale agrees much better.

Even if $\tilde t_{\rm e}$ is treated as an arbitrary parameter
(corresponding to horizontal translation in Fig.9) the fit is not
entirely satisfactory: both tails of the distribution are not well
described.  For this reason we have also tried the empirical
distribution $P(\tilde t) = (1+a\tilde t)^{-b}$, where $a$, $b$ are
adjustable constants.  The result is shown
as a solid line in Fig.9.  It is quite satisfactory.
Incidentally, the data curve which is most deviant at large $\tilde t$
has the smallest value of $\tilde E$, where the effect of the finite
cutoff in $t$ in the numerical integrations is likely to be most
serious.  The best fitting parameters from all sets of data are given
in Table 1.  The quality of the fit is comparable to that in Fig.9
for all cases except $W_0 = 3$ and uniform initial conditions.  Also
shown in this Table is the time $\tilde t_{1/2}$ at which $P = 1/2$.
This illustrates the trends mentioned in connection with eqs.(9) and
(12), for which $\tilde t_{1/2} \simeq 0.092$:  the result is smaller
for larger $W_0$ and uniform initial conditions.



For the non-escapers the interpretation of the numerical results is less
quantitative.  The fact that the majority of non-escapers move
initially on retrograde orbits is consistent with the known stability
of retrograde motions (\S3.2.1).  The approximate limit of $\tilde E = 
0.3$ (for the planar problem) is approximately in line with Fig.4.
Nevertheless this Figure refers to the three-dimensional problem, and
we also find that significant numbers of non-escapers exhibit prograde 
motion.  An example is shown in Fig.10.  Note that this is a cross
section in {\sl configuration}
space, i.e. it is not a Poincar\'e surface of section.  (Actually there 
is also a tiny fraction of stable prograde orbits even in the 
planar problem, as one can 
just discern in Fig.5.)  The sharply defined outer boundary of the
plot in Fig.10
 suggests a regular orbit, and is a very common feature of such
plots for non-escapers, both prograde and retrograde.

These features contrast sharply with those for escapers.  Fig.11
illustrates a similar plot for one of the longest lasting escapers.
Note the absence of sharp boundaries, suggesting an effectively
stochastic orbit.  Further study showed that this orbit spends a long
time in a restricted part of phase space, corresponding to the denser
part of the distribution.  At early and late times, however, its
distribution was wider.


\bigskip\noindent{\bf  5. Discussion and Conclusions}




\smallskip

It was mentioned in the introduction that this was our motivation for
studying the distribution of escape times.  The main numerical
results were given in Sec.2 as a function of the energy of a star.  In
a model star cluster, however, the primordial escapers have a range of
energies, and we have carried out similar calculations for this
distribution.  For a King model with $W_0=3$ we found
that the fraction of stars which were ``primordial escapers'' was
$0.137\pm0.003$ (90\% confidence),
and that their median time of escape was $1360\pm180$ $N$-body units.

Now let us compare this with the lifetime of a simulation.  Here we
take the initial parameters of the ``collaborative experiment''
(Heggie \etal 1998), in which the initial mass function was of the
Salpeter form, i.e. $f(m){\rm d}m\propto m^{-2.35}{\rm d}m$, with a range of
masses which scales to $0.1M_\odot < m < 1.5M_\odot$.  The time $t_{\rm h}$
for half of the mass to escape beyond the tidal radius is given in
Table 2, for one of the participating groups.  We also give the
fraction $F$ of all stars which were primordial escapers and which
would have remained within the tidal radius at $t_{\rm h}$, assuming that
the potential had not evolved.

\bigskip

From these results we see that the time scale on which the primordial
escapers leave the cluster is at least comparable with the time scale on which
it loses mass.  It is this process which
makes scaling of these results problematic.  The time scale for loss
of primordial escapers is
approximately independent of $N$, whereas the two-body relaxation time
scale is approximately proportional to $N$ (except for the Coulomb
logarithm).  If  the time scale of escape had been very short, it
might have been possible to create appropriate initial conditions by
running the model for a short time, during which the escapers leave
but the effects of relaxation are still minor.

In some $N$-body calculations (e.g. Portegies Zwart \etal 1998) the
tidal effect is treated as a {\sl cutoff}, i.e. stars are removed as
soon as they cross the radius $r_{\rm t}$, though the equations of motion
ignore all but the first term in eq.(2).  This procedure avoids the
difficulty of setting up self-consistent initial conditions.  By a
calculation analogous to that in Sec.3.1, the time scale for escape of
phase volume across the tidal boundary changes to 
$$
t_{\rm e}={\pi\over8}{\sqrt{-2\ecrit}r_{\rm t}\over E-\ecrit},
$$
where now $\ecrit=- GM/r_{\rm t}$.  The important change is in the
dependence on $(E-\ecrit)$: stars with energy just above the escape
energy (which may result from two-body encounters in a cluster, for
example) escape much more readily with a tidal cutoff than with a
tidal field.  On the other hand these models also do not contain any
primordial escapers, and so the benefit of a short  escape  time scale
is not so necessary for scaling purposes.

To return to the case of a tidal field, it is clear that scaling the
results of $N$-body simulations requires rather detailed modelling of
the escape process, and relevant data for such a task are presented
in this paper.  On the other hand we have considered a simpler
problem, as the potential is fixed.  In an $N$-body simulation the
potential alters as the cluster loses mass.  In addition, the
primordial escapers are supplemented by stars which gain enough energy
to escape by two-body encounters.  The theory of relaxation implies
that the energy which they have at this point depends on $N$, and
therefore the determination of the time scale on which they escape is
a more complicated problem.

Since relaxation is a diffusive process, it can be argued that the
energy which a star reaches before escaping, if the time taken is
$t_{\rm e}$, is given by an expression which scales as $E - E_{\rm crit} \sim
E_{\rm crit} (t_{\rm e}/t_{\rm r})^{1/2}$, where $t_{\rm r}$ is the relaxation time.  Since
we also have 
$$
t_{\rm e}\sim {1\over\omega}\left({E_{\rm crit}\over E - E_{\rm crit}}\right)^2,
$$
by eqs.(3), (4) and (9), it follows that $t_{\rm e}\sim\sqrt{t_{\rm r}/\omega}$,
i.e. the typical escape time varies as the geometric mean of the
crossing and relaxation times.

Besides the collaborative experiment to which we have referred, there
are many simulations in the literature whose interpretation may be
complicated by the presence of primordial escapers, and which have usually
gone unnoticed.  For example Wielen (1975) found that the
mass-dependence of the escape rate on stellar mass was much weaker
than expected theoretically; this could be expected if a substantial
number of escapers was primordial, though other factors referred to by
Wielen (such as mass segregation) surely also play a role.  Sugimoto
\& Makino (1989) studied the evolution of binary clusters by placing
two concentrated King models in a circular relative orbit, and drew
attention to the importance of escape for driving the clusters
together.  The results of Fukushige \& Heggie (1995) on the dissolution of
tidally limited
clusters by mass loss from stellar evolution, and those by Vesperini
\& Heggie (1997) on the effect of dnamical evolution on the mass
function, are also subject to the
effects of primordial escapers.  

When the escape time scales found in Secs.2 and 5.1 are scaled to real
clusters, it is found that the escape time scales can be surprisingly
long.  For example, in a King model with $W_0 = 3$ at a galactocentric
distance $R_{\rm g}$ in a galaxy with circular speed $V$, the median escape
time is about $\displaystyle{6\times10^9\left({R_{\rm g}\over10{\rm
kpc}}\right)\left({220{\rm km/sec}\over V}\right)}$ yr.  Furthermore, we have seen that it is
possible for stars to remain inside the tidal radius indefinitely,
even with energies above the energy of escape.  Now in two clusters a
few stars have been observed with radial velocities surprisingly close
to the escape velocity of a dynamical model (Gunn \& Griffin 1979,
Meylan, Dubath \& Mayor 1991).  Usually interpreted as ejecta from energetic
interactions in the core (Davies, Benz \& Hills 1994, Sigurdsson \& Phinney 1995), or
perhaps as evidence of a centrally condensed population of dark
remnants (Larson 1984), it is possible that they are simply trapped
within the cluster with energies above the escape energy.

It must be remembered that our
results apply to an idealised problem, and in particular to a cluster
on a circular galactic orbit.  On the other hand Johnston {\sl et al} (1999)
have pointed out that clusters may be accompanied by an {\sl
extratidal} population of bound stars even in the elliptic case, and so the
existence of such stars within the tidal radius is perfectly
plausible.  It is also not hard to see how such orbits can be
populated:  at any stage a cluster will contain stars on retrograde
orbits at energies below the energy of escape;  as the cluster loses
mass their energy will increase, placing them in retrograde orbits
within the cluster but above the escape energy.

\bigskip

\noindent{\bf Acknowledgements}

T.F. acknowledges financial support from the ``Research for the
Future'' Program of the Japan Society for the Promotion of Science,
grant no. JSPS-RFTP 97P01102, and thanks Jun Makino for helpful
discussions.

\bigskip

{\parindent=-0.5truein
\leftskip=0.5truein

{\bf References}

Anosova J.P., Orlov V.V., 1994, CeMDA, 59, 327

Baumgardt H., 1997, PhD Thesis, University of Heidelberg (in German)

Benest D., 1974, A\&A, 32, 39

Brunini A., 1996, in Muzzio J.C., Ferraz-Mello S., Henrard J., eds,
Chaos in Gravitational $N$-Body Systems.  Kluwer, Dordrecht, p.79

Chandrasekhar S., 1943, ApJ, 97, 54

Chernoff D.F., Weinberg M.D., 1991, in Janes K., ed, ASP
Conf. Ser. 13, The Formation and
Evolution of Star Clusters.
ASP, San Francisco, p.373

Davies M.B., Benz W., Hills J.G., 1994, ApJ, 424, 870

de la Fuente Marcos R., 1995, A\&A, 301, 407

Fukushige T., Heggie D.C., 1995, MNRAS, 276, 206

Gunn J.E., Griffin R.G., 1979, AJ, 84, 752

Hayli A., 1970, A\&A, 7, 17

Heggie D.C., Giersz M., Spurzem R., Takahashi K., 1998, in Andersen
J., ed, Highlights of Astron., Vol.11B.  Reidel, Dordrecht, p.591

Heggie D.C., Mathieu R.D., 1986, in Hut P., McMillan S., eds, LNP267, The Use
of Supercomputers in Stellar Dynamics.  Springer-Verlag,
Berlin, p.233

H\'enon M., 1969, A\&A, 1, 223

H\'enon M., 1970, A\&A, 9, 24.

Huang T.-Y., Innanen K.A., 1983, AJ, 88, 1537

Jefferys W.H., 1976, AJ, 81, 983

Johnston K., Sigurdsson S., Hernquist L., 1999, MNRAS, 302, 771

Johnstone D., 1993, AJ, 105, 155

Kandrup H.E., Siopis C., Contopoulos G., Dvorak R., 1999,
astro-ph/9904046

Keenan D.W., Innanen K.A., 1975, AJ, 80, 290

King I.R., 1960, AJ, 65, 122

King I.R., 1966, AJ, 71, 64

Kwast T., 1977, Post. Astron, 25, 105

Larson R.B., 1984, MNRAS, 210, 763

Lee H.-M.,  Goodman J., 1995, ApJ, 443, L109

Lee H.-M., Ostriker J.P., 1987, ApJ, 322, 123

MacKay R.S., 1990, Phys. Lett. A, 145, 425

Makino J., Aarseth S.J., 1992, PASJ, 44, 141

Meylan G., Dubath P., Mayor M., 1991, ApJ, 383, 587

Murison M.A., 1989, AJ, 98, 2346 and 2383

Pietrovskaia I.V., 1977, Post. Astron., 25, 59

Portegies Zwart S.F., Hut P., Makino J., McMillan S.L.W., 1998, 
A\&A, 337, 363

Press W.H., Teukolsky S.A., Vetterling W.T., Flannery B.P., 1992,
Numerical Recipes, 2nd ed.  Cambridge Univ. Press, Cambridge

Ross D.J., Mennim A., Heggie D.C., 1997, MNRAS, 284, 811

Sigurdsson S., Phinney E.S., 1995, ApJS, 99, 609

Spitzer L., Jr, 1987, Dynamical Evolution of Globular Clusters.
Princeton Univ. Press, Princeton

Spitzer L., Jr, Shull J.M., 1975, ApJ, 201, 773

Sugimoto D., Makino J., 1989, PASJ, 41, 1117

Takahashi K., Portegies Zwart S.F., 1998, ApJ, 503, L49 

Vesperini E., Heggie D.C., 1997, MNRAS, 289, 898


Wielen R., 1975, in Hayli A., ed, Proc. IAU
Symp. 69, Dynamics of Stellar Systems. D. Reidel, Dordrecht, p.119

Wiggins S., 1992, Chaotic Transport in Dynamical Systems.
Springer-Verlag, New York

}

\bigskip\noindent
{\bf Appendix:  Motion near a Lagrangian Point}

\medskip
Adding the Coriolis acceleration (the cross product in eq.(2)) to the
acceleration derived from eq.(5), and choosing units in which
$\omega = 1$, we have the linearised equations of motion
$$\eqalign{
\ddot x - 2\dot y - 9x &= 0\cr
\ddot y + 2\dot x + 3y &= 0.\cr
}
$$
Their general solution is 
$$
{x\choose y} = A {-\mu \choose4-\sqrt{7}}\exp \mu t +
B{\mu\choose4-\sqrt{7}}\exp[ -\mu t] + C{-\nu\cos (\nu t+\theta)\choose
(4+\sqrt{7})\sin (\nu t+\theta)},
$$
where $\mu =  \sqrt{1+2\sqrt{7}}$, $\nu = \sqrt{2\sqrt{7}-1}$ and $A$,
$B$, $C$ and $\theta$ are arbitrary constants.  The energy is 
$E = E_{\rm crit} + C^2(10\sqrt{7}+49) + AB(196-40\sqrt{7})$.

When $A = B = C$ we have equilibrium at the Lagrange point; when only
$C$ is non-zero we have a linear approximation to the periodic orbit
referred to in Sec.3.2.2; when also $B$ is non-zero we have a local
approximation to an orbit which approaches the periodic orbit
asymptotically.

\bigskip

\noindent{\bf Tables}

\medskip

$$\vbox{
\centerline{Table 1}
\medskip
\centerline{Empirical fit to distribution of $\tilde t$}
\medskip
$$\vbox{
\settabs
\+$W_0$	&\quad Non-uniform	&\quad$5.98$	&\quad$0.869$	&\quad$0.680$\cr
{\hrule}
\smallskip
\+$W_0$	&Initial 	&$a$	&$b$		&$\tilde t_{1/2}$\cr
\+	&conditions	&	&		&	\cr
\smallskip
{\hrule}
\smallskip
\+3	&Non-uniform	&$1.80$	&$0.869$	&$0.680$\cr
\+3	&Uniform	&$7.68$	&$0.579$	&$0.301$\cr
\+7	&Non-uniform	&$3.96$	&$0.758$	&$0.377$\cr
\+7	&Uniform	&$8.31$	&$0.729$	&$0.191$\cr
\+12	&Uniform	&$5.98$	&$0.914$	&$0.189$\cr
\smallskip
{\hrule}
}$$
}$$

\bigskip
$$\vbox{
\centerline{Table 2}
\medskip
\centerline{Half-Mass Time for one set of N-body Results of the Collaborative
Experiment}
$$\vbox{
\settabs \+
$t_{\rm h}$\quad\quad\quad	&$0.093\pm0.005$\quad	&\quad$0.087\pm0.005$
&\quad$0.080\pm0.005$	&\quad$0.073\pm0.003$	\cr
\hrule
\smallskip
\+$N$	&$4096$	&$8192$	&$16384$	&$32768$	\cr
\+$t_{\rm h}$	&$260$	&$410$	&$620$		&$1000$		\cr
\+$F$	&$0.093\pm0.005$	&$0.087\pm0.005$
&$0.080\pm0.005$	&$0.073\pm0.003$	\cr
\smallskip
\hrule
}$$
Note: the times are given in $N$-body units (Heggie \& Mathieu 1986),
as in Section 2.
}$$

\parindent=0truein
\leftskip=0truein

{\bf Figure Captions}
\medskip

Fig.1  Creation of a cluster model with a tidal radius of 2 (in
suitably scaled units).  The solid
curve shows the potential well of an isolated model, and stars are
created with energies up to the solid horizontal line.  Then the
galactic tide (long-dashed) is added.  This creates a tidal radius at the
maximum of the combined potential (lower dashed curve), but the stars
have energies up to the upper dashed curve.  Many lie above the escape
energy (corresponding to the value of the combined potential at the tidal radius.)

\smallskip

Fig.2  Equipotentials of the effective gravitational potential $\phi$ experienced by a star
in a tidally limited cluster, including the inertial force caused by
the rotation of the frame of reference about the galactic centre.
The cluster potential is approximated by that of a point mass, and
units are such that $\phi = -1/\sqrt{x^2+y^2} - 3x^2/2$ in the plane
$z=0$.  The Lagrange points are the saddle points.

\smallskip

Fig.3  Cumulative distribution of escape times from a King model
with $W_0 = 3$.  $P$ is the probability (i.e. fraction of initial
conditions) that the lifetime exceeds $t$.  The initial conditions are 
selected non-uniformly, as explained in the text.  The relative excess 
energy is, from left to right, $\tilde E = 0.24$, $0.16$,
$0.12$, $0.08$, $0.06$, $0.04$ and $0.03$.

\smallskip

Fig.4  Fraction of initial conditions with ``infinite'' escape time,
i.e. ``non-escapers'', if the potential is fixed.  The initial
conditions stated in the key are described in Sec.2.1.

\smallskip

Fig.5.  Surface of section for motion in the $x,y$ plane at energy
$E_{\rm crit}$.  Each point corresponds to the value of $x$ and $\dot x$
at the instant when the orbit  crosses the $y$-axis with $\dot y>0$.
Different orbits are characterised by different symbols.

\smallskip

Fig.6.  The orbit asymptotic (as $t\to\infty$) to the Lagrangian
point $x =r_{\rm t}$ (in units with $G = M = \omega = 1$).

\smallskip

Fig.7.  An orbit asymptotic to a periodic orbit (at the right of the
figure), at some energy above
$E_{\rm crit}$.

\smallskip

Fig.8.  A set of orbits asymptotic to the periodic orbit, all at one energy
above $E_{\rm crit}$.  The projection of the phase-space onto the $x$, $y$ plane is shown.

\smallskip

Fig.9 The distribution of the scaled escape time $t_s$ (denoted by
$\tilde t$ in eq.[11])
for escapers only, in the case of non-uniform initial conditions in a
King potential with $W_0 = 3$ (cf. Fig.3).  Results are shown as
dashed curves for the same values of the scaled energy excess $\tilde
E$ as in Fig.3.  Also shown  are the graph of
eq.(12), i.e. the long-dashed curve,  with $t_{\rm e}$ determined as stated in the text, and the
empirical fit (solid curve) with parameters from Table 1.

\smallskip

Fig.10  A non-escaper in the potential of a King model with $W_0 =
3$.  The orbit is shown every time it crosses the plane $y=0$
in one direction, by plotting a point with coordinates $x$
(horizontally) and $z$.  Orbits in $x<0$  are prograde.

\smallskip

Fig.11  As Fig.10, but for a late escaper.  Here $\tilde E = 0.04$
and the escape time was 37445.

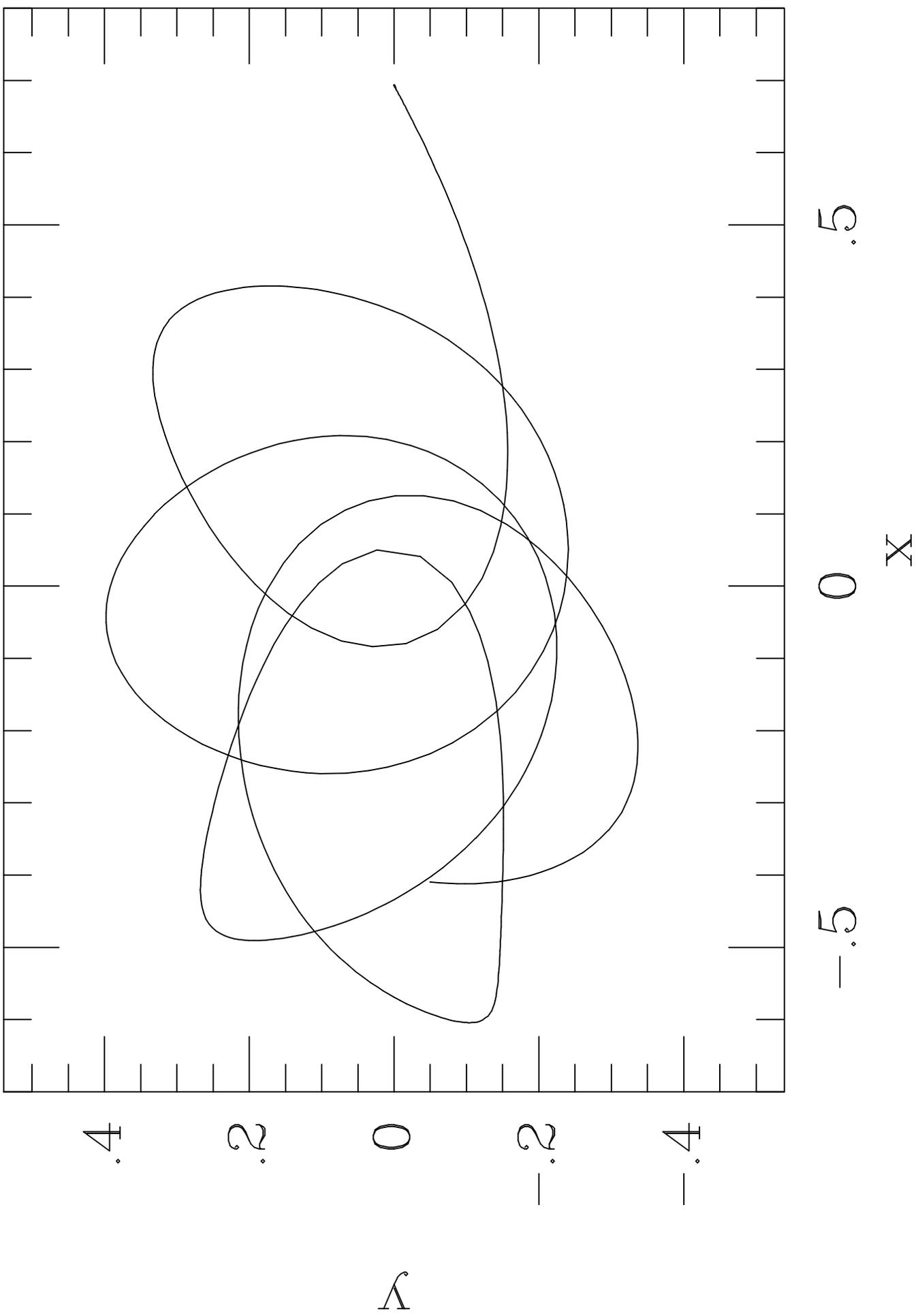
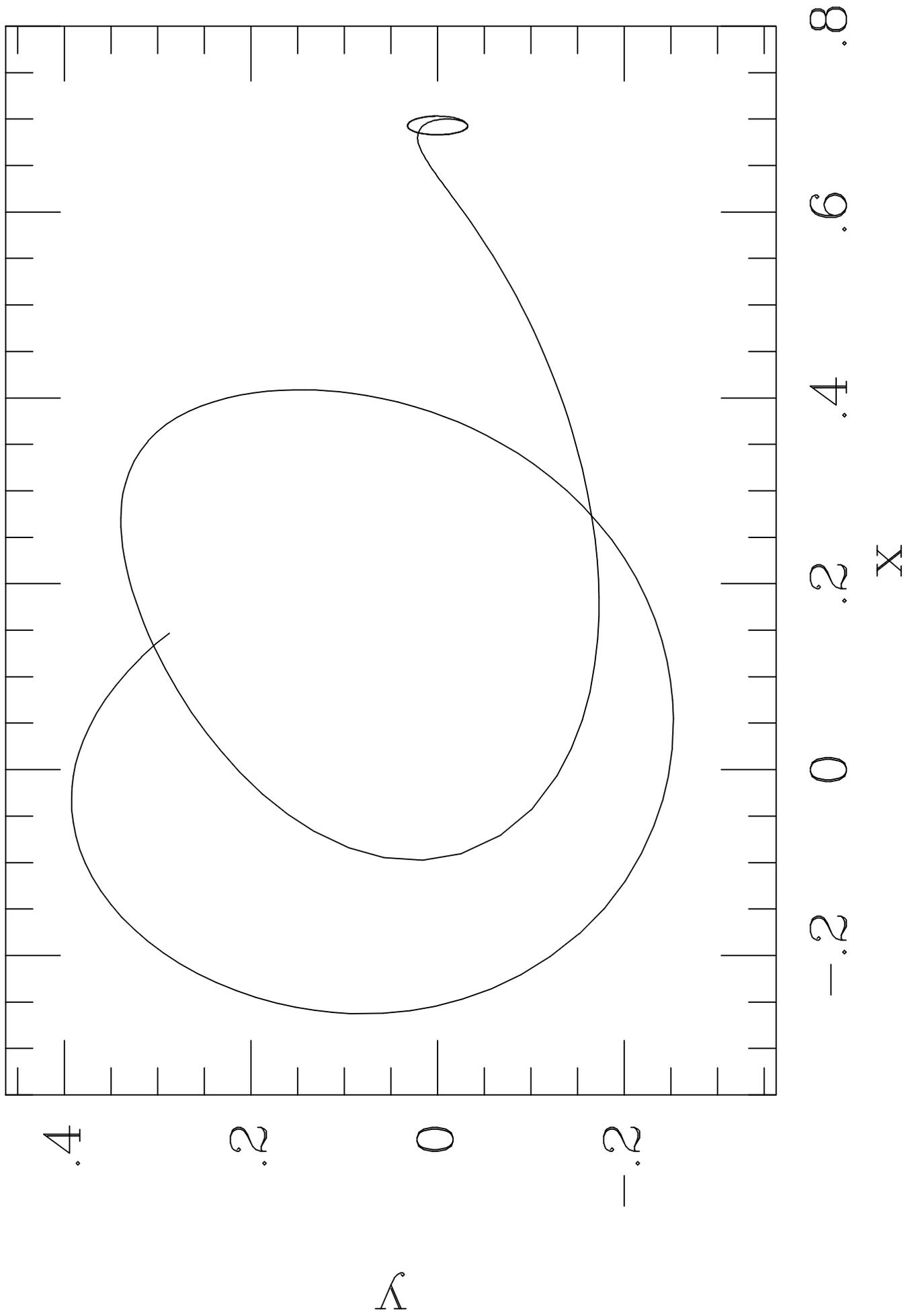
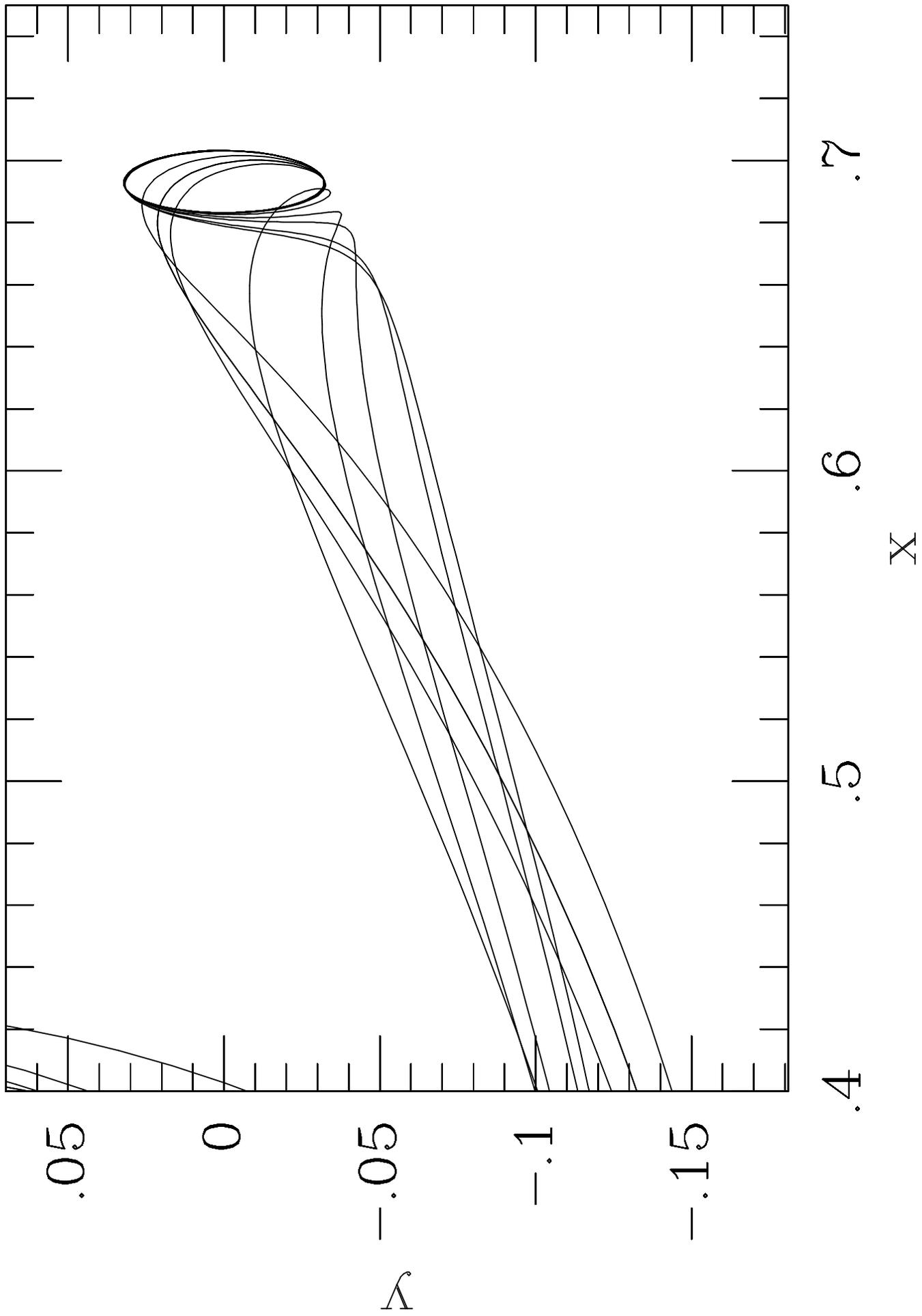
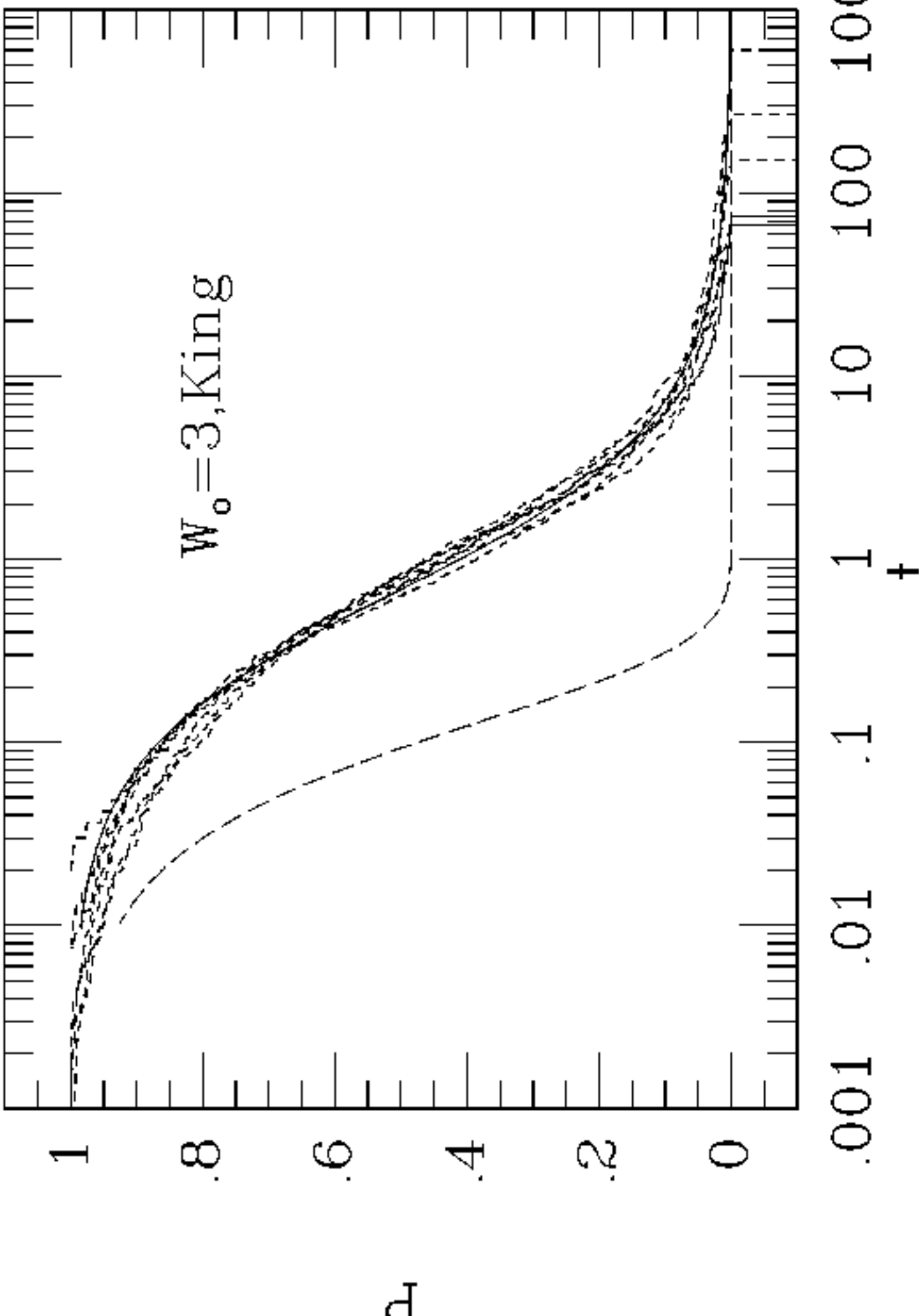
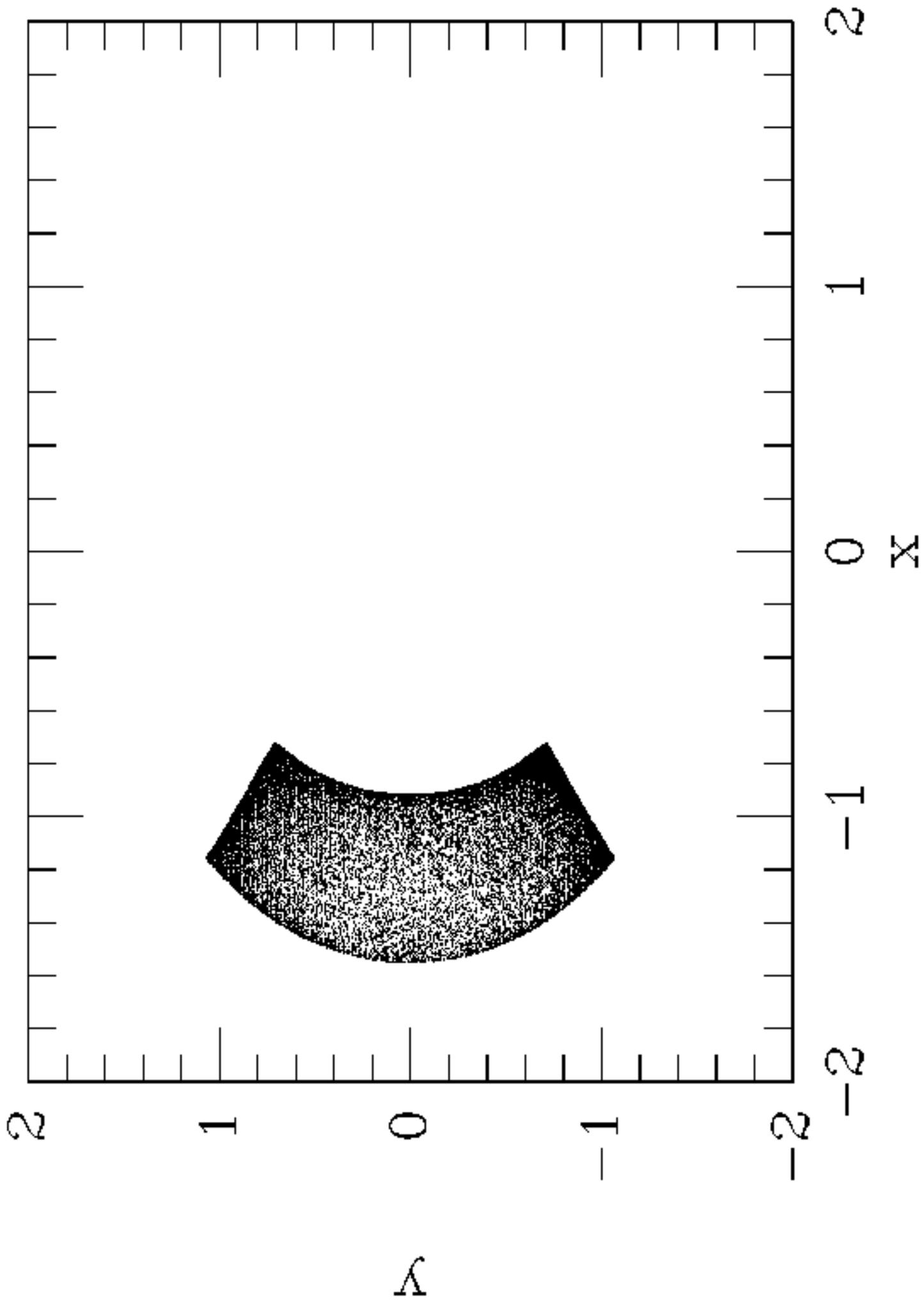
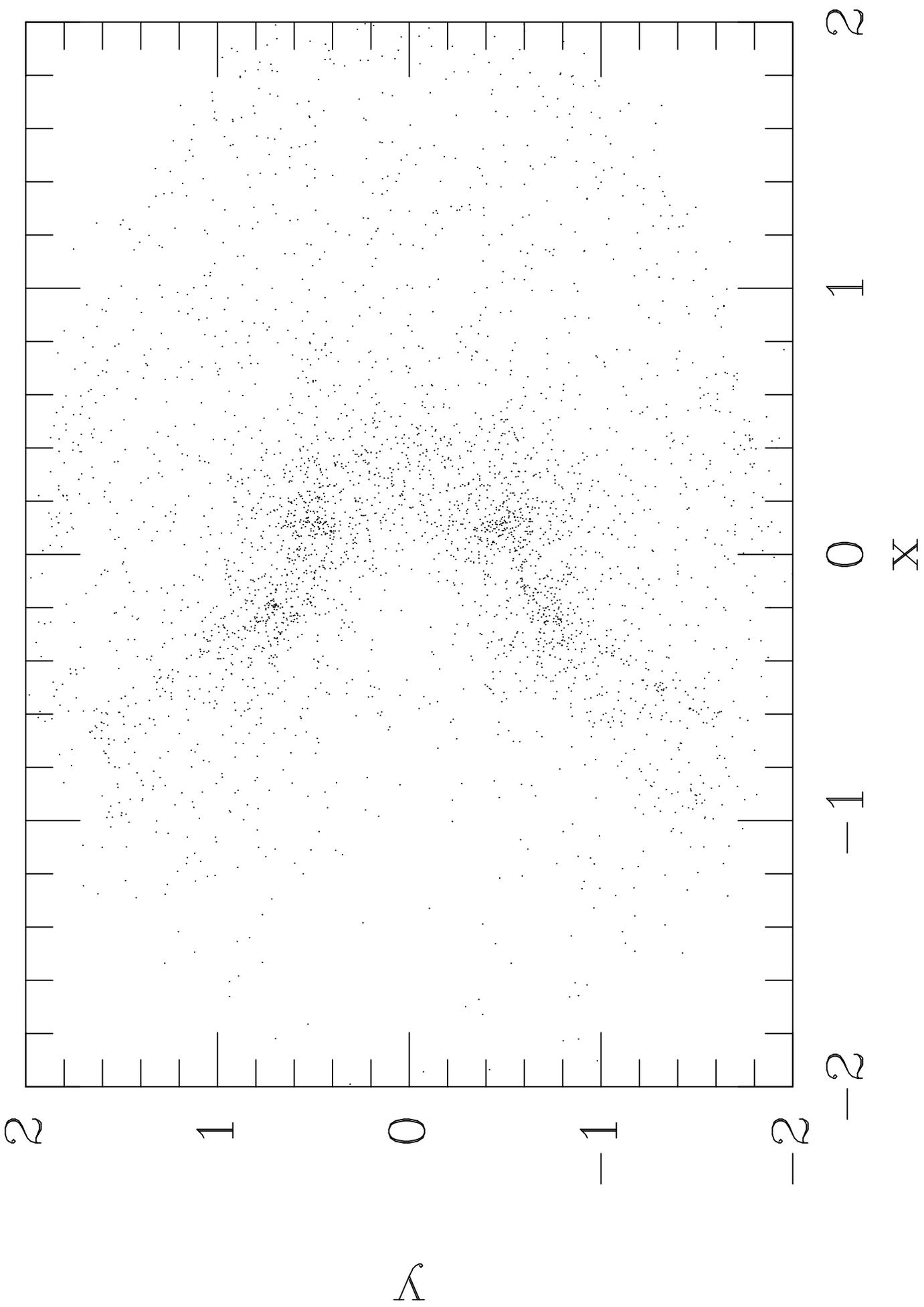
\bye